\documentclass[runningheads]{llncs}
\usepackage[T1]{fontenc}
\usepackage{graphicx}
\usepackage{booktabs}
\usepackage{tabularx}
\usepackage{longtable}
\usepackage{mathtools}
\usepackage[ruled,vlined]{algorithm2e}
\usepackage{makecell}

\usepackage{comment}
\usepackage{diagbox}
\usepackage{enumitem} 
\usepackage[justification=centering]{caption}

\usepackage{amsmath,amsfonts}
\usepackage{color,xcolor,soul}
\usepackage{geometry}

\newcommand{\hlc}[2][yellow]{ {\sethlcolor{#1} \hl{#2}} }

\begin{document}
\title{Elementary Constructions of Best Known Quantum Codes  \thanks{This research was supported by Kenyon College Summer Science Scholars research program. N. Aydin is the corresponding author.}}
\newgeometry{left=3.5cm, right=3.5cm, top=2.4cm, bottom=1.9cm}
\titlerunning{Elementary Constructions of Quantum Codes}
%
\author{Nuh Aydin\inst{1}\orcidID{0000-0002-5618-2427} \and
Trang T. T. Nguyen\inst{1} \and
Long B. Tran\inst{1}}
\authorrunning{N. Aydin et al.}
%
\institute{Kenyon College, Gambier OH 43022, USA 
\email{\{aydinn,nguyen7,tran2\}@kenyon.edu}}
%
\maketitle              
\begin{abstract}
Recently, many good quantum codes over various finite fields $\mathbb{F}_q$ have been constructed from codes over extension rings  or mixed alphabet rings via some version of a Gray map. We show that most of these codes can  be obtained more directly from cyclic codes or their generalizations over $\mathbb{F}_q$. Unless explicit benefits are demonstrated for the indirect approach, we believe that direct and more elementary methods should be preferred. 
\end{abstract}
\section{Introduction and Motivation}\label{sec1}

Building large scale quantum computers  is a very active area of research currently. A critical part of this effort is the ability to control quantum errors. The idea of quantum error-correcting codes was first introduced in 1990s \cite{Gottesman}, \cite{CSS}. It is well known that quantum error-correcting codes (QECCs) can be constructed from classical codes with certain properties.  
 The construction method given in \cite{QECC4} is called the CSS construction and it has been extensively used in the literature. There is a large body of literature that makes use of the CSS construction.  Since then researchers have
investigated various methods of using classical error correcting codes to construct new QECCs. Recently, many  quantum codes over finite fields $\mathbb{F}_q$ with good parameters have been constructed from codes over extension rings of $\mathbb{F}_q$ or mixed alphabet rings via some version of a Gray map. We observe that most of these codes can be directly constructed from codes over $\mathbb{F}_q$, without needing indirect way of using codes over extension rings and taking their Gray images. We adhere to the principle that a simple explanation or construction is more desirable than a complicated one unless there is a specific benefit or advantage to the complicated method.

This paper focuses on more direct ways of constructing quantum codes from classical codes, including cyclic, constacyclic, quasi-cyclic (QC), quasi-twisted (QT), and polycyclic codes using the CSS construction, Hermitian construction, and Lisonek-Singh methods. Cyclic codes and their generalizations are among the most important classes of codes in algebraic coding theory. Our work shows once more the usefulness of these  codes. We have reconstructed many codes without using extension rings or mixed alphabet rings, and in some cases, we discovered new codes. To construct a quantum code over  $\mathbb{F}_q$, we only use codes over $\mathbb{F}_q$, and occasionally codes over $\mathbb{F}_{q^2}$. We do not use any extension rings or Gray maps. 

The material in this paper is organized as follows. In section two we recall some basic
definitions; in section three we give a summary of search methods that we have used to construct quantum error correcting codes, and the last two sections present the codes that we have recreated or improved upon.
\section{Basic Definitions}
 Let $\mathbb{F}_q$ denote the finite field of order $q$ where $q$ is a prime power. This field is also denoted by $GF(q)$, the Galois field of order $q$. A code $C$ of length $n$ over $\mathbb{F}_q$ is a subset of $\mathbb{F}_q^n$. Elements of $C$ are called codewords. If $C$ is a vector subspace of $\mathbb{F}_q^n$, then it is called a linear code. In practice, virtually all codes are linear. The minimum Hamming weight (distance) of a linear code is defined as $d=\min\{w_H(c): 0\not = c \in C \}$, where $w_H(c)=\#\{i: c_i\not = 0 \}$ denotes the Hamming weight of a vector $c=(c_0,c_1,\dots,c_{n-1})$. 
 Any linear code has three fundamental parameters: the length $n$, the dimension $k$, and the minimum distance $d$. Such a code over $\mathbb{F}_q$ is referred to as an $[n,k,d]_q$ code.  One of the most important and challenging problems in coding theory is to determine the optimal values of the parameters of a linear code and to explicitly construct codes whose parameters  attain optimal values, or come as close to them as possible. There is an online database of best known linear codes (BKLC) \cite{website}. To use the database, one first chooses the size of the finite field (the alphabet) $q=2,3,4,5,7,8,9$, then the length $n$ (within the bounds for each alphabet), and the dimension $k$.  The database then reports the best available theoretical  upper bound on the minimum distance $d$, and the minimum distance of a BKLC for this length and dimension over the chosen alphabet. When the two values coincide, this means that optimal codes have been found for these parameters. When there is a gap between the two values, it means that codes with better parameters could potentially be found, even though their existence cannot be guaranteed from this information alone. One can observe that in most cases, optimal codes are not known. They are generally known when either $k$ or $n-k$ is relatively small.

 Researchers often use computer searches to find codes with better parameters than currently BKLCs. However, exhaustive searches over  all linear codes are not feasible due to two fundamental facts.  First, computing the minimum distance of an arbitrary linear code is NP-hard \cite{NP-Hard}, and it quickly becomes infeasible for larger dimensions. Second, for a given length, dimension, and  the finite field $GF(q)$, the number of linear codes is very large, so large that exhaustive computer searches are not feasible for most lengths and dimensions. Given these inherent computational complexity challenges, researchers often focus on promising subclasses of linear codes with rich mathematical structures. Cyclic codes and their various generalizations play a key role in algebraic coding theory.

\begin{definition}
A linear code $ C $ of length $ n $ over a finite field $ \mathbb{F}_q $ is called cyclic  if for every codeword $ v = (v_0, v_1, \ldots, v_{n-1}) \in C $, the word obtained by cyclically shifting $ v $ by one position to the right, that is, $ \sigma(v) = (v_{n-1}, v_0, v_1, \ldots, v_{n-2}) $, is also in $ C $. 
\end{definition}

Cyclic codes have a central place in coding theory. They establish a key link between coding theory and algebra. If we represent a vector $ v = (v_0, v_1, \ldots, v_{n-1}) \in \mathbb{F}_q^n$ as a polynomial $v(x)=v_0+v_1x+\cdots+v_{n-1}x^{n-1}$, then its cyclic shift  $\sigma(v)$ corresponds to  $x\cdot v(x) \mod (x^n-1)$. It follows that cyclic codes are ideals in the quotient ring  $\mathbb{F}_q[x]/\langle x^n-1 \rangle$, which is a principal ideal ring. For every non-trivial cyclic code $C$, there is a non-zero polynomial $g(x)$ of least degree in $C$ that divides $x^n-1$ and generates  $C$, i.e., $C=\langle g(x) \rangle=\{p(x)g(x) \mod x^n-1: p(x) \in \mathbb{F}_q[x]  \}$. This special polynomial $g(x)$ is called the standard generator polynomial of $C$ and it divides any other generator of $C$. There is a one-to-one correspondence between the divisors of $x^n-1$ in  $\mathbb{F}_q[x]$ and cyclic codes of length $n$ over  $\mathbb{F}_q$. Therefore, all cyclic codes of length $n$ over  $\mathbb{F}_q$ can be obtained from the factorization of $x^n-1$ into irreducibles. Let $C$ be a cyclic code with the standard generator $g(x)$, and    let $x^n-1=g(x)h(x)$. Then $h(x)$ is called the check polynomial of $C$ and it characterizes codewords by the following condition.  $$ v(x) \in \mathbb{F}_q[x] \text{  is a codeword if and only if }  v(x)h(x)=0 \mod x^n-1. $$ 

\noindent Cyclic codes have a number of useful generalizations. One of them is the class of constacyclic codes.
\begin{definition}
A linear code $ C $ of length $ n $ over a finite field $ \mathbb{F}_q $ is called constacyclic if there exists a nonzero element $ \lambda $ in $ \mathbb{F}_q $ such that for every codeword $ v = (v_0, v_1, \ldots, v_{n-1}) \in C $, its constacyclic shift $ \sigma_{\lambda}(v) = (\lambda v_{n-1}, v_0, v_1, \ldots, v_{n-2}) $ is also in $ C $. 
\end{definition}

Note that when $\lambda=1$, a constacyclic code is a cyclic code.  Therefore, cyclic codes are a special case of constacyclic codes. The algebraic structure of a constacyclic code is very similar to that of a cyclic code. Constacyclic codes are ideals in the quotient ring $\mathbb{F}_q[x]/\langle x^n-\lambda \rangle$ and they can be obtained from the factorization of   $x^n-\lambda$ into irreducibles over $\mathbb{F}_q$. 

Quasi-cyclic (QC) codes are another generalization of cyclic codes. Those linear codes which are invariant under a cyclic shift by some fixed number of positions are called quasi-cyclic.

\begin{definition}
A linear code $ C $ of length $ n $ over a finite field $ \mathbb{F}_q $ is called a quasi-cyclic code of index $ \ell $ if for every codeword $ v = (v_0, v_1, \ldots, v_{n-1}) \in C $, the word obtained by cyclically shifting $ v $ by $ \ell $ positions, that is, $ \sigma^{\ell}(v) = (v_{n-\ell}, v_{n-\ell+1}, \ldots, v_{n-1}, v_0, v_1, \ldots, v_{n-\ell-1}) $, is also in $ C $. 
\end{definition}

We can also generalize QC codes in the same way that constacyclic codes generalize cyclic codes. This gives us the class of quasi-twisted (QT) codes.

\begin{definition}
A linear code $ C $ of length $ n $ over a finite field $ \mathbb{F}_q $ is called a quasi-twisted (QT)  code of index $ \ell$ and shift constant $\lambda$ if for every codeword $ v = (v_0, v_1, \ldots, v_{n-1}) \in C $, the word obtained by the constacyclic shift of $v$ by $ \ell$ positions is also a codeword, i.e.,  $\sigma_{\lambda}^{\ell}(v) = (\lambda v_{n-\ell}, \lambda v_{n-\ell+1}, \ldots, \lambda v_{n-1},  v_0, v_1, \ldots, v_{n-\ell-1})$, is also in $ C $. 
\end{definition}

A QC code is a QT code with shift constant $\lambda=1$. The length of a QT code is of the form $n=m\ell$. Algebraically, a QT code is an $R$-submodule of $R^{\ell}$ where  $R=\mathbb{F}_q[x]/\langle x^n-\lambda \rangle$.  Many record-breaking linear codes have been found from the class of QC and QT codes by computer searches. For example,  27  new linear codes that are QT or QC with better parameters than the previous BKLCs in the database \cite{website} were found in \cite{Aydin2} (and there are many other papers in the literature that present new codes from the class of QC and QT codes). A large number of BKLCs in the database \cite{website} are QT codes.  Yet another generalization of cyclic codes are polycyclic codes that have received much attention recently.

\begin{definition} \cite{quasipoly}
     A linear code $C$ of length $n$ over $\mathbb{F}_q$ is said to be polycyclic with respect to $\mathbf{v} = (v_0, v_1, \ldots, v_{n-1}) \in \mathbb{F}_q^n$ if for any codeword $(c_0, c_1, \ldots, c_{n-1}) \in C$, its right polycyclic shift, $(0, c_0, c_1, \ldots, c_{n-2}) + c_{n-1}(v_0, v_1, \ldots, v_{n-1})$ is also a codeword. Similarly, $C$ is left polycyclic with respect to $\mathbf{v} = (v_0, v_1, \ldots, v_{n-1}) \in \mathbb{F}_q^n$ if for any codeword $(c_0, c_1, \ldots, c_{n-1}) \in C$, its left polycyclic shift $(c_1, c_2, \ldots, c_{n-1}, 0) + c_0(v_0, v_1, \ldots, v_{n-1})$ is also a codeword. If $C$ is both left and right polycyclic, then it is bi-polycyclic. 
\end{definition}
Note that if the vector $\mathbf{v}$, also called the associate vector of the polycyclic code, is $(1,0, \ldots, 0)$ then a polycylic code is an ordinary cyclic code, and when $\mathbf{v}=(\lambda,0, \ldots, 0)$, the polycyclic code is a constacyclic code.  We will work only with right polycyclic codes, which we refer to them as simply polycyclic codes from now. Algebraically, polycyclic codes are ideals in the quotient ring $\mathbb{F}_q[x]/\langle f(x) \rangle$ where $f(x)=x^n-v(x)$ and $v(x)$ is the associate vector in polynomial form. For every polycyclic code $C$, there is a standard generator polynomial $g(x)$ such that $C=\langle g(x) \rangle$ and $g(x)|f(x)$. The polynomials $f(x)$ and $g(x)$ are listed in Table 5 below that lists  good QECCs from polycyclic codes. Polycyclic codes can be generalized in a way that QC codes generalize cyclic codes. 


\begin{definition}\cite{quasipoly}
A linear code \( C \) is said to be an \( r \)-generator quasi-polycyclic (QP) code of index \( \ell \) if it has a generator matrix of the form
\[
\begin{pmatrix}
G_{11} & G_{12} & \cdots & G_{1\ell} \\
G_{21} & G_{22} & \cdots & G_{2\ell} \\
\vdots & \vdots & \ddots & \vdots \\
G_{r1} & G_{r2} & \cdots & G_{r\ell}
\end{pmatrix}
\]
where each \( G_{ij} \) is a generator matrix of a polycyclic code. The special case of a 1-generator quasi-polycyclic code with \( \ell \) blocks has a generator matrix of the form
\[
\begin{pmatrix}
G_{11} & G_{12} & \cdots & G_{1\ell}.
\end{pmatrix} 
\]
\end{definition}

Note that when each $G_{ij}$ is a generator matrix of a constacyclic code then we obtain QT codes as a special case. Finally, we have a further generalization. A generalized quasi-polycyclic module $\mathcal{P}$ is an $\mathbb{F}_q[x]$-module of the form $\mathcal{P}=\prod_{i=1}^{r}\mathbb{F}_q[x]/\left\langle f_i(x)\right\rangle,$
where  $f_1(x), f_2(x),\ldots, f_r(x)$ are monic polynomials in $\mathbb{F}_q[x]$ \cite{sigmaPoly}. 
\begin{definition}[Generalized quasi-polycyclic code \cite{sigmaPoly}]
A generalized quasi-polycyclic (GQP) code of length $n=n_1+\cdots + n_r$ is an $\mathbb{F}_q[x]$-submodule of a generalized polycyclic module $\mathcal{P}=\displaystyle\prod_{i=1}^{r}\mathbb{F}_q[x]/\langle f_i(x)\rangle$, such that  for all $i\in\{1,\ldots,r\}$, $n_{i}= \text{deg}~f_i(x)$. 
\end{definition}

When all of the polynomials $f_i(x)$ in the definition of GQP codes are of the form $x^{n_i}-a_i$ for some non-zero element $a_i$ of $\mathbb{F}_q$, then we obtain the class of multi-twisted (MT) codes as a special case \cite{MT}.

\begin{definition}[1-generator generalized quasi-polycyclic code \cite{sigmaPoly}]
A GQP code $\mathcal{C}$ over $\mathcal{P}$ is said to be  a 1-generator GQP code  if there exists an $\underline{a}(x)=\left(a_1(x), a_2(x), \ldots, a_r(x)\right) \in \mathcal{P}$ such that
$$
\mathcal{C}=\mathbb{F}_q[x] \underline{a}(x)=\langle\underline{a}(x)\rangle=\left\{g(x) \underline{a}(x) \mid g(x) \in \mathbb{F}_q[x]\right\}.
$$
If $\mathcal{C}=\langle\underline{a}(x)\rangle$, then the monic polynomial $h(x)$ of minimum degree satisfying
$
h(x) \underline{a}(x)=\mathbf{0}
$ 
is called the check polynomial of $\mathcal{C}$.
\end{definition}

\begin{definition} \cite{QECC4}
A quantum error-correcting code $ Q $, denoted by $((n, K, d))_q$ or $[[n, k, d]]_q$, is a subspace of the $ n $-fold tensor product $ (\mathbb{C}^q)^{\otimes n} $ of the complex vector space $ \mathbb{C}^q $. The code has dimension $\dim(Q) = K = q^k$ and minimum distance $ d $, i.e., any error acting on at most $ d - 1 $ of the tensor factors can be detected or has no effect on the code. 
\end{definition}
Similar to the case of classical codes, a main problem in the field of QECCs is to construct codes with best possible parameters. There are databases of best known quantum codes that are available online \cite{database}, \cite{databasePaper}, \cite{website}, \cite{yves}.

\section{Methods of Constructing QECCs from Classical Codes}
In this section, we survey and summarize a few methods of constructing quantum error-correcting codes (QECCs) from classical codes. CSS construction and the Lisonek-Singh method are among the most direct approaches for constructing  QECCs. Both methods effectively translate classical coding techniques into the quantum domain, facilitating the creation of robust QECCs. First, we recall some relevant basic facts about classical codes.

For every \( u = (u_1, \ldots, u_n) \) and \( v = (v_1, \ldots, v_n) \) in \( \mathbb{F}_q^n \), the (Euclidean) inner product \( u \cdot v \) is defined as
\[ u \cdot v = u_1 v_1 + u_2 v_2 + \cdots + u_n v_n. \]
The dual code of \( C \) is defined as
\[ C^\perp = \{ v \in \mathbb{F}_2^n : u \cdot v = 0 \, \forall u \in C \}. \]
A code that is contained in its dual, \( C \subseteq C^\perp \), \( C \) is called self-orthogonal or weakly self-dual. A code $C$ that contains its dual, \( C^\perp  \subseteq C\), is called dual-containing. If \( C = C^\perp \) then we say that \( C \) is self-dual. If $C \cap C^\perp =\{ 0\}$ then $C$ (and $C^\perp$) is called  a linear complementary dual (LCD) code.

\subsection{CSS Construction}
We can apply the  CSS construction to obtain QECCs  from cyclic, constacyclic, QC, QT and polycyclic codes. For this, we need two codes such that one is contained in the dual of the other one. Hence for cyclic and constacyclic codes,  this condition is characterized by the ideal inclusion
\begin{align*}
    \langle g(x) \rangle \supseteq \langle g(x)f(x) \rangle.
\end{align*}
So we take $C_2^{\perp}=\langle g(x)f(x)\rangle \subseteq \langle g(x) \rangle =C_1 $. 

\begin{theorem}[CSS construction \cite{QECC4}]
Let  $C_1$ and $C_2$ be two linear codes over $\mathbb{F}_q$ with parameters $[n,k_1,d_1]_q$ and $[n,k_2,d_2]_q$ with $C_2^{\perp}\subseteq C_1.$ Then there exists a QECC with parameters $[[n, k_1 + k_2 - n, \min(d_1, d_2)]]_q$. In case $C_1$ is a dual-containing code, that is, $C_1^{\perp} \subseteq C_1$, there exists a QECC with parameters $[[n, 2k_1 - n, d_1]]_q$.
 \end{theorem}

Here is an example  that illustrates the construction of a quantum code from a dual-containing cyclic code.

\begin{example}
    Let $q=11$ and $n=19$. Consider the factorization of $x^{19}-1$ in $\mathbb{F}_{11}[x]$:
\begin{align*}
 x^{19}-1=(x + 10)(x^3 + x^2 + 2x + 10)(x^3 + 2x^2 + 8x + 10)(x^3 + 3x^2 + 6x + 10)\\(x^3 + 3x^2 + 9x + 10)(x^3 + 5x^2 + 8x + 10)(x^3 + 9x^2 + 10x + 10)
\end{align*}

    Let $f(x)=x^3+9x^2+10x+10$, one of the irreducible factors of $x^{19}-1$ over $\mathbb{F}_{11}$. We find, with the help of Magma software \cite{magmaOnline}, that $C=\langle f(x)\rangle$ is a dual-containing cyclic code with parameters $[19,16,3]_{11}$. The dual code has the parameters $[19,3,16]_{11}$. We then obtain the QECC with parameters $[[19,13,3]]_{11}$ applying the CSS construction. Note that this code has the same parameters as the best known quantum code given in our database \cite{database} .
\end{example}
\subsection{Lisonek-Singh Construction - Quantum Construction X}

This is a generalized version of Construction X using Hermitian codes \cite{marcolla_roggero}, first introduced in \cite{lisonek}.  The method refines and generalizes the construction of QECCs from nearly self-orthogonal codes in \cite{Ezerman}. The process begins with a nearly Hermitian self-orthogonal linear code over $\mathbb{F}_4$ and utilizes the information about the code’s Hermitian hull to extend its length, ensuring that the extended code remains self-orthogonal, which then can create QECCs code using the CSS construction. This approach can be generalized to $\mathbb{F}_{q^2}$ linear classical codes to create QECCs using cyclic, QC or QT codes.


\subsubsection{Background Notation}

For codes $ C \subseteq \mathbb{F}_{q}^{2n} $ of even length, we split a vector  $ v = (a|b) $ with $ a, b \in \mathbb{F}_{q}^n $ and define the symplectic weight as
\[
\mathrm{swt}(a|b) = |\{i \mid i \in \{1, \ldots, n\}, (a_i, b_i) \neq (0, 0)\}|.
\]
For a set $ S \subseteq \mathbb{F}_2^{2n} $, we denote the symplectic minimum distance by $ d_S(S) = \min\{\mathrm{swt}(x - y) \mid x, y \in S, x \neq y\} $.

\subsubsection{Inner products and dual codes}
For finite fields whose degree of extension is even, we define the Hermitian inner product on $ \mathbb{F}_{q^2}^n $ as
\[
\langle u, v \rangle_H := \sum_{i=0}^{n-1} u_i v_i^q, \tag{II.4}
\]
where $ u = (u_0, u_1, \ldots, u_{n-1}) $ and $ v = (v_0, v_1, \ldots, v_{n-1}) $ are vectors in $ \mathbb{F}_{q^2}^n $.

The Hermitian dual $ C^{\perp_H} $ is given by
\[
C^{\perp_H} := \{ b \in \mathbb{F}_{q^2}^n \mid \langle c, b \rangle_H = 0, \forall c \in C \}. \tag{II.5}
\]

The Hermitian hull of $ C $ is defined to be $ C \cap C^{\perp_H} $. The code $ C $ is Hermitian self-orthogonal if it is equal to its Hermitian hull, or equivalently, if $ C \subseteq C^{\perp_H} $, i.e., $ C $ is contained in its Hermitian dual. A code $C$ is called nearly self-orthogonal if $\dim(C^{\perp_H})-\dim(C^{\perp_H} \cap C)$ is a small positive integer.

\subsubsection{Stabilizer codes/ Hermitian construction}
Let $C \subseteq \mathbb{F}_{q^2}^n$ be an $\mathbb{F}_{q^2}$-linear code with parameters $[n, k]_{q^2}$. Define $e := k - \dim(C \cap C^{\perp_H})$. Then there exists an $[[n + e, n - 2k + e, d(Q)]]_q$ quantum stabilizer code $Q$ with

\[
\begin{aligned}
    d(Q) & \ge \min \left(\operatorname{wgt}(C^{\perp_H} \setminus (C \cap C^{\perp_H})), \operatorname{wgt}((C + C^{\perp_H}) \setminus C)\right) + 1\\
    & \ge \min\left\{d(C^{\perp_H}), d(C + C^{\perp_H}) + 1\right\} \\
    \text{and} \\
    d(Q) & \le \operatorname{wgt}(C^{\perp_H} \setminus (C \cap C^{\perp_H})).
\end{aligned}
\]

\begin{example} 

    Consider the following polynomials in $\mathbb{F}_4[x]$:
\begin{align*}
   g_{1}(x) &= x^{16} + x^{15} + x^{14} + \omega^2 x^{11} + \omega x^{10} + \omega^2 x^8 + \omega^2 x^7 + \omega x^6 + \omega x^5 + \omega^2 x^4 + \omega x^2 \\
    g_{2}(x) &= \omega^2 x^{16} + x^{15} + \omega^2 x^{14} + \omega^2 x^{13} + \omega x^9 + \omega x^8 + \omega x^7 + x^6 + \omega^2 x^4 + \omega^2 x^3 + x^2 
\end{align*}

\noindent where $\omega$ is a primitive element of $\mathbb{F}_4$. The two-generator QT code with index $\ell = 2$, co-index $m = 15$, and $\lambda = \omega^2$, generated by $g_1$ and $g_2$, is a $[30,15,8]_4$ code $C$. The Hermitian dual $C^{\perp_H}$ also has parameters $[30,15,8]_4$. Since $C = C^{\perp_H}$, $C$ is a self-dual code. Therefore, the Hermitian hull is $C \cap C^{\perp_H}$ with parameters $[30,0,30]_4$, and $C + C^{\perp_H}$ has parameters $[30,30,1]_4$. This implies that $\text{wgt}(C^{\perp_H} \setminus (C \cap C^{\perp_H})) = d(C^{\perp_H}) = 8$ and $\text{wgt}((C + C^{\perp_H}) \setminus C) = 8 > d(C + C^{\perp_H}) = 2$. As stated in \cite{Ezerman}, the improved lower bound need not be tight. With a suitable choice of the complement of $C \cap C^{\perp_H}$ in $C$, we can achieve $d(Q) = 8$. With $e = k - \dim(C \cap C^{\perp_H}) = 15$, we obtain a $[[n + e, n - 2k + e, d(Q)]]_q$ quantum code, specifically a $[[45,15,8]]_4$ quantum code.

\end{example}

\section{Computational Results}


As  mentioned at the beginning, our goal is to match, and if possible, improve on the parameters of best known QECCs through some of the direct constructions. Thus, we begin with CSS construction using cyclic codes. If we are unable to replicate the existing codes in the literature, we then proceed with CSS construction using constacyclic, QC, and QT codes, concluding with construction X. To push the boundaries even further, we also use  polycyclic and generalized polycyclic codes as ingredients in the CSS construction, aiming to achieve not only best-known quantum codes but also potentially record-breaking parameters directly.

Most of our codes (both improved and tied compared to best-known quantum codes) are obtained from cyclic codes over $\mathbb{F}_q$ via the  CSS construction. One of the main reasons for using the cyclic CSS construction is its simplicity. Unlike more complex methods, cyclic codes over $\mathbb{F}_q$  do not require complicated algebraic techniques. This simplicity makes it an attractive choice, especially for initial explorations and for constructing codes that are easy to analyze and implement.

Despite its simplicity, the cyclic CSS construction yielded many quantum codes that are competitive with the best-known codes. In some cases, these codes have better parameters, while in others, they match the performance of the best-known codes. This combination of simplicity and effectiveness makes the cyclic CSS method a powerful tool in the field of quantum error correction.

\subsection{Explanation of Tables}

The tables below display the codes we have obtained through direct constructions, along with their generators. Quantum codes without an asterisk have been replicated from the papers in the literature (that is, codes with the same parameters have been presented in the literature) where the reported QECCs were obtained from codes over an extension ring. Those with an asterisk indicate improvements we have achieved, either  on $d$ or $k$. We consider a quantum code with parameters $[[n,k,d]]_q$ better than a quantum code with parameters $[[n,k',d']]_q$   if either $k=k' \text{ and }d> d'$ or  $ k>k' \text{ and } d=d'$.

In the tables below, a polynomial is represented as a list containing only coefficients to save space. The ordering is such that coefficients of the highest degree term are in the left-most position. For instance, the polynomial $x^3 + 2x + 3x + 1$ is represented as $1231$. 
We also use some letters to represent numbers with two digits, with A = 10 and B = 11. For non-prime fields, we use $w$ to represent a root of the irreducible polynomial used to define the extension field. For $GF(9)$, $w$ is a root of $x^2 + 2x + 2$, and for $GF(25)$, it is a root of $x^2 + 2x + 4.$

Table 1 shows the QECCs from CSS construction using cyclic codes over non-binary fields. Table 2 is for the binary codes which is a very important special case in coding theory. Table 3 presents CSS construction using QC and QT codes. Table 4 is based on construction X. Finally, we also obtained some good (the same parameters as the codes reported in the literature or databases)  and new  quantum codes (codes whose parameters do not appear in the literature or a database) from polycyclic and generalized quasi-polycyclic codes (GQPs). They are listed in Tables 5 and 6. It is quite likely that many other quantum codes  could have been obtained using these direct methods.  In this work, we present a sample of such codes.

\begin{table}[ht]
\begin{center}
\begin{minipage}{\textwidth}
\caption{Non-binary QECCs constructed from cyclic codes}\label{tab1}
\begin{tabular*}{\textwidth}{@{\extracolsep{\fill}}llll@{\extracolsep{\fill}}}
\toprule
$[[n,k,d]]_q$ & Generator Polynomials & Parameters &{References}\\ 
\midrule
\hline
$[[6,2,3]]_7$ &$156$ & $[6, 4, 3]_7$ & \cite{YunGao}\\
\hline
$[[10,6,3]]_{11}$ &$166$ & $[10,8,3]_{11}$ &\cite{YunGao}\\
\hline
$[[11,1,6]]_{11}$ &$1 6 A15A$ & $[11,6,6]_{11}$ &\cite{Liu2}\\
\hline
$[[12,10,2]]_{5}*$ &$1285$ & $[12,11,2]_{5}$ &\cite{Prakash}\\
\hline
$[[12,6,4]]_{13}$ &$17114$ & $[12,9,4]_{13}$ &\cite{Prakash}\\
\hline
$[[12,4,5]]_{13}$ &$13$ & $[12,8,5]_{13}$ &\cite{Verma}\\
\hline
$[[14,8,3]]_{7}$ &$1661$ & $[14,11,3]_{7}$ &\cite{Liu1}\\
\hline
$[[16,8,2]]_5$&$10003$ & $[16,12,2]_{5}$ &\cite{ashraf2016}\\
\hline
$[[16,12,2]]_{13}$&$105$ & $[16,14,2]_{13}$ &\cite{Dinh2020}\\
\hline
$[[16,14,2]]_{13}*$&$108$ & $[16,15,2]_{13}$ &\cite{Dinh2020}\\
\hline
$[[16,10,3]]_{25}$&$1w^{21}w^{21}3$ & $[16,13,3]_{25}$ &\cite{Prakash}\\
\hline
$[[16,10,4]]_{17}$&$164B$ & $[16,13,4]_{17}$ &\cite{Prakash}\\
\hline
$[[18,16,2]]_7*$&$15$ & $[18,17,2]_{7}$ &\cite{pandey}\\
\hline
$[[18,10,3]]_7*$&$15016$ & $[18,14,3]_{7}$ &\cite{Prakash}\\
\hline
$[[20,18,2]]_5*$&$14$ & $[20,19,2]_{5}$ &\cite{pandey}\\
\hline
$[[20,14,3]]_{11}$&$1951$ & $[20,17,3]_{11}$ &\cite{pandey}\\
\hline
$[[24,22,2]]_{9}*$&$1w^7$ & $[24,23,2]_{9}$ &\cite{pandey}\\
\hline
$[[24,18,3]]_{9}*$&$1w22$ & $[24,21,3]_{9}$ &\cite{pandey}\\
\hline
$[[30,28,2]]_{7}*$&$15$ & $[30,29,2]_{7}$ &\cite{pandey}\\
\hline
$[[30,20,3]]_{7}*$&$165361$ & $[30,25,3]_{7}$ &\cite{pandey}\\
\hline
$[[33,25,3]]_{11}$&$12321$ & $[33,29,3]_{11}$ &\cite{YunGao}\\
\hline
$[[36,30,2]]_3$ &$1212$ & $[36,33,2]_3$  &\cite{YunGao}\\
\hline
$[[36,32,2]]_7*$ &$104$ & $[36,34,2]_7$  &\cite{pandey}\\
\hline
$[[36,8,5]]_7*$ &$104201305104305$ & $[36,22,5]_7$  &\cite{pandey}\\
\hline
$[[36,20,3]]_7*$ &$104000305$ & $[36,8,3]_7$  &\cite{pandey}\\
\hline
$[[40,24,2]]_5$ & $102040301$ & $[40,32,2]_5$  &\cite{ashraf2016}\\
\hline
$[[40,36,2]]_5*$ & $102$ & $[40,38,2]_5$  &\cite{ashraf2016}\\
\hline
$[[40,32,3]]_5$& $13312$ & $[40,36,3]_5$ &\cite{pandey}\\
\hline
$[[40,18,4]]_5*$& $10001$ & $[40,36,2]_5$ &\cite{ashraf2016}\\
\hline
$[[60,48,2]]_5$& $1444144$ & $[60,54,2]_5$ &\cite{ashraf2016}\\
\hline
$[[60,48,3]]_5*$ &$1031444$ & $[60,54,3]_5$ &\cite{ashraf2016}\\
\hline
$[[60,56,2]]_5*$&$124$ & $[60,58,2]_5$ &\cite{ashraf2016}\\
\hline
$[[72,68,2]]_3$&$122$ & $[72,70,2]_3$ &\cite{Dinh2020}\\
\hline
$[[80,64,2]]_5$&$100040004$ & $[80,72,2]_5$ &\cite{ashraf2016}\\
\hline
$[[80,76,2]]_5$&$103$ & $[80,78,2]_5$ &\cite{Dinh2020}\\
\hline
$[[88,48,2]]_5$&$100030004000400010001$ & $[88,68,2]_5$ &\cite{ashraf2016}\\
\hline
$[[88,68,2]]_5*$&$10201020302$ & $[88,78,2]_5$&\cite{ashraf2016}\\
\hline
$[[90,86,2]]_5$&$141$ & $[90,88,2]_5$ &\cite{Dinh2020}\\
\hline
$[[96,80,2]]_5$&$100040002$ & $[96,88,2]_5$ &\cite{ashraf2016}\\
\hline
$[[96,88,2]]_5*$&$10402$ & $[96,92,2]_5$ &\cite{ashraf2016}\\
\hline
$[[100,92,2]]_5$&$11111$ & $[100,96,2]_5$ &\cite{ashraf2016}\\
\hline
$[[100,98,2]]_5*$&$14$ & $[100,99,2]_5$ &\cite{ashraf2016}\\
\hline
$[[112,64,2]]_5$&$1000300040002000100030004$ & $[112,88,2]_5$ &\cite{ashraf2016}\\
\hline
$[[112,88,2]]_5$&$1030403020402$ & $[112,100,2]_5$ &\cite{ashraf2016}\\
\hline
$[[120,104,2]]_5$&$1130002303$ & $[120,112,2]_5$ &\cite{ashraf2016}\\
\hline
$[[120,112,2]]_5$&$130014$ & $[120,116,2]_5$ &\cite{ashraf2016}\\
\hline
$[[120,116,2]]_5*$&$142$ & $[120,118,2]_5$ &\cite{ashraf2016}\\
\hline
$[[120,108,3]]_5*$&$1234324$ & $[120,114,3]_5$ &\cite{pandey}\\
\hline
$[[168,164,2]]_7*$& $164$ & $[168,166,2]_7$ &\cite{Dinh2020}\\
\hline
\end{tabular*}
\end{minipage}
\end{center}
*: better code compared to the codes in references.
\end{table}

\begin{table}[ht]

This table presents binary quantum codes obtained from cyclic codes over $GF(2)$.

\begin{center}
\begin{minipage}{\textwidth}
\caption{Quantum binary codes}\label{tab6}
\begin{tabular*}{\textwidth}{@{\extracolsep{\fill}}lllll@{}}
\toprule
$[[n,k,d]]_q$ & Generator Polynomials & Parameters & Compared codes & References\\ 
\midrule
\hline
$[[42,18,3]]_2$& $1010100010001$ &[42, 30, 3]& $[[42, 10,3]]_2$& \cite{Z4}\\
\hline
$[[90,34,5]]_2$& $11111011111000000000000011111$ &$[90, 62, 5]$& $[[90,24,5]]_2$& \cite{Z4}\\
\hline
$[[[28, 26, 2]]_2$& $11$ &$[28, 27, 2]$& $[[[28, 26, 2]]_2$& \cite{website}\\
\hline
$[[15,7,3]]_2$& $11001$ &$[15,11,3]$& $[[15,7,3]]_2$& \cite{Lv}\\
\hline
$[[35,29,2]]_2$& $101110010111001011100101110010111$ &$[35,3,20]$& $[[35,29,2]]_2$& \cite{website}\\
\hline

\end{tabular*}
\end{minipage}
\end{center}
\end{table}

\begin{table}[ht]
\begin{center}
\begin{minipage}{\textwidth}
\caption{QECCs constructed from QC and QT codes}\label{tab2}
\begin{tabular*}{\textwidth}{@{\extracolsep{\fill}}llll@{\extracolsep{\fill}}}
\toprule
$[[n,k,d]]_q$ &Old codes&{References}\\ 
\midrule
\hline
$[[60,56,2]]_5*$ &$[[60,54,2]]_5$&\cite{YunGao}\\
\hline
$[[72,64,2]]_3$ &$[[72,64,2]]_3$&\cite{YunGao}\\
\hline
$[[96,92,2]]_5*$ &$[[96,90,2]]_5$&\cite{YunGao}\\
\hline
$[[112,108,2]]_5$ &$[[112,108,2]]_5$&\cite{Dinh2020}\\
\hline
$[[112,116,2]]_3$ &$[[112,116,2]]_3$&\cite{Dinh2020}\\
\hline
$[[120,116,2]]_5*$ &$[[120,114,2]]_5$&\cite{YunGao}\\
\hline
$[[140,112,2]]_5$ &$[[140,136,2]]_5$&\cite{ashraf2016}\\
\hline

\end{tabular*}
\end{minipage}
\end{center}
*: better code compared to the codes in references.
\end{table}

\begin{table}[ht]
\begin{center}
\begin{minipage}{\textwidth}
\caption{QECCs constructed from Lisonek-Singh construction}\label{tab3}
\begin{tabular*}{\textwidth}{@{\extracolsep{\fill}}llll@{\extracolsep{\fill}}}
\toprule
$[[n,k,d]]_q$ &{References}\\ 
\midrule
\hline
$[[48,6,11]]_4$ &\cite{Ezerman}\\
\hline
$[[24,8,4]]_9*$ &\cite{Prakash}\\
\hline
$[[6,0,4]]_4$ &\cite{Dastbasteh}\\
\hline
$[[16,4,6]]_{25}$ &\cite{Prakash}\\
\hline
$[[16,8,4]]_{25}$ &\cite{Prakash}\\
\hline
\end{tabular*}
\end{minipage}
\end{center}
*: better code compared to the codes in references.
\end{table}


\newcolumntype{T}{>{\centering\arraybackslash\tiny}p{5.3cm}}
\newcolumntype{S}{>{\centering\arraybackslash\tiny}p{4cm}}
\newcolumntype{L}{>{\centering\arraybackslash\tiny}p{3cm}}
\newcolumntype{V}{>{\centering\arraybackslash\tiny}p{1.5cm}}

\begin{table}[ht]

\begin{center}
\begin{minipage}{\textwidth}
\caption{Good and new quantum codes from polycyclic codes }\label{tab4}

\begin{tabular*}{\textwidth}{@{\extracolsep{\fill}}lTllll@{}}
\toprule
$[[n,k,d]]_q$ & $f$ & $g$ & Parameters & Compared codes & References\\ 
\midrule
\hline
$[[12, 10, 2]]_5*$ & $1012432043044$ & $12$ & $[12, 11, 2]_5$ & $[[12, 9, 2]]_5$& \cite{pandey}\\
\hline
$[[20, 18, 2]]_5*$ & $140300442223221134102$ & $11$ & $[20, 19, 2]_5$ & $[[20, 15, 2]]_5$&\cite{pandey}\\
\hline
$[[18, 16, 2]]_7*$ & $1206440404343305511$ & $13$ & $[18, 17, 2]_7$ & $[[18, 15, 2]]_7$&\cite{pandey}\\
\hline
$[[30,28,2]]_7*$ & $1240336055361430431021404450005$ & $12$ & $[20,29,2]_7$ &$[[30,25,2]]_7$& \cite{pandey}\\
\hline
$[[32,28,2]]_7$ & $163335431201401365401215124154035$ & $116$ & $[32,30,2]_7$ &$[[30,28,2]]_7$& \cite{pandey}\\
\hline
$[[36,32,2]]_7*$ & $1554623464062532555266640354165240165$ & $104$ & $[36,34,2]_7$ &$[[36,30,2]]_7$& \cite{pandey}\\
\hline
$[[24,22,2]]_9*$ & $15546w46w0625325w552w^26664w41652$ & $104$ & $[24,23,2]_9$ &$[[24,21,2]]_9$& \cite{pandey}\\
\hline
\end{tabular*}
\end{minipage}
\end{center}
*: better code compared to the codes in references.
\end{table}


\newgeometry{left=1cm, right=1cm, top=2cm, bottom=2cm}

\renewcommand{\arraystretch}{2}
\begin{table}[ht]
\begin{center}
\begin{minipage}{\textwidth}
\caption{QECCs constructed from generalized quasi-polycyclic codes}\label{tab5}
\begin{tabular*}{\textwidth}{@{\extracolsep{\fill}}| V|T|S|V|V|V|V|@{\extracolsep{\fill}}}
\toprule
$[[n,k,d]]_q$ & $(f_1;f_2)$ &$(g_1;g_2)$ &$(p_1;p_2)$ & Parameters &Compared codes & References\\ 
\midrule

\hline
$[[7,5,2]]_7$ & 
\makecell[{{p{5cm}}}]{\raggedright$f_{1}=130521, n_1=5$ \\$f_{2}=115, n_2=2$} & 
\makecell[{{p{5cm}}}]{\raggedright$g_{1}=14323$ \\$g_{2}=12$} & 
\makecell[{{p{5cm}}}]{\raggedright$p_{1}=3$ \\$p_{2}=2$} & 
$[7,1,7]_7$ & $[[7,5,2]]_7$ & \cite{tushar}\\

\hline
$[[12,10,2]]_7$ & 
\makecell[{{p{5cm}}}]{\raggedright$f_{1}=15365311, n_1=7$ \\$f_{2}=122512, n_2=5$} & 
\makecell[{{p{5cm}}}]{\raggedright$g_{1}=1161364$ \\$g_{2}=14226$} & 
\makecell[{{p{5cm}}}]{\raggedright$p_{1}=6$ \\$p_{2}=1$} & 
$[12,1,12]_7$ & $[[12,10,2]]_7$& \cite{chen} \\

\hline
$[[15,13,2]]_7$ & 
\makecell[{{p{5cm}}}]{\raggedright$f_{1}=106432011, n_1=8$ \\$f_{2}=13030012,n_2=7$} & 
\makecell[{{p{5cm}}}]{\raggedright$g_{1}=13234254$ \\$g_{2}=1261321$} & 
\makecell[{{p{5cm}}}]{\raggedright$p_{1}=2$ \\$p_{2}=2$} & 
$[15,1,15]_7$ & $[[15,13,2]]_7$& \cite{yves}\\

\hline
$[[27,25,2]]_7$ & 
\makecell[{{p{5cm}}}]{\raggedright$f_{1}=15412, n_1=4$ \\$f_{2}=111235660262561653224641, n_2=23$} & 
\makecell[{{p{5cm}}}]{\raggedright$g_{1}=1435$ \\$g_{2}=16314361621424266221351$} & 
\makecell[{{p{5cm}}}]{\raggedright$p_{1}=4$ \\$p_{2}=4$} & 
$[27,1,27]_7$ & $[[27,25,2]]_7$& \cite{yves} \\

\hline

$[[32,28,2]]_7$ & 
\makecell[{{p{5cm}}}]{\raggedright$f_{1}=160011150, n_1=8$\\$f_{2}=1420522535566006612456406, n_2=24$} & 
\makecell[{{p{5cm}}}]{\raggedright$g_{1}=1025631$ \\$g_{2}=16313412663412336005132$} & 
\makecell[{{p{5cm}}}]{\raggedright$p_{1}=55$ \\$p_{2}=20$} & 
$[32,2,25]_7$ & $[[32,28,2]]_7$& \cite{dinh2020} \\

\hline
$[[40,36,2]]_5$ & 
\makecell[{{p{5cm}}}]{\raggedright$f_{1}=144003114, n_1=8 $ \\ $f_{2}= 110333404313221010032223440431320, n_2=32$} & 
\makecell[{{p{5cm}}}]{\raggedright$g_{1}=1442043$ \\$g_{2}=12013310214210323140$\\ $1302102103$} & 
\makecell[{{p{5cm}}}]{\raggedright$p_{1}=1$ \\$p_{2}=34$} & 
$[40,2,30]_5$ & $[[40,36,2]]_5$& \cite{dinh2020} \\

\hline
$[[60,56,2]]_5*$ & 
\makecell[{{p{5cm}}}]{\raggedright$f_{1}=13103142043, n_1=10$ \\$f_{2}=113200230333204422411422220111303304003 $\\ $043001301331, n_2=50$} & 
\makecell[{{p{5cm}}}]{\raggedright$g_{1}=113002043$\\$g_2=120123441412344203214320331$\\$024040210141443410303$} & 
\makecell[{{p{5cm}}}]{\raggedright$p_{1}=20$ \\$p_{2}=42$} & 
$[60,2,45]_5$ & $[[60,54,2]]_5$& \cite{dinh2020-2} \\

\hline
$[[63,59,2]]_7*$ & 
\makecell[{{p{5cm}}}]{\raggedright$f_{1}=1 1 1 4 1 2 4 0 1 1 2 1,n_1=11$ \\$f_{2}=1 1 2 0 1 1 0 0 4 1 2 4 2 3 1 0 2 4 4 1 0 4 2 1 4 0 2 1 1$\\$ 4 2 1 2 4 1 0 1 4 1 2 3 3 2 3 4 1 2 1 2 1 4 1 4, n_2=52$} & 
\makecell[{{p{5cm}}}]{\raggedright$g_{1}=1 3 1 1 3 0 1 3 1 4$ \\$g_{2}=1 4 3 4 1 0 4 2 0 3 2 0 1 2 2 3 1 1 2 3$\\$ 2 0 2 3 2 2 0 1 0 0 0 1 3 1 0 0 0 1 0 1 0 4 4 4 3 1 0 1 0 4 1$} & 
\makecell[{{p{5cm}}}]{\raggedright$p_{1}=62$ \\$p_{2}=3$} & 
$[63,2,52]_7$ & $[[63,57,2]]_7$& \cite{kongbo}\\

\hline
$[[75,71,2]]_5*$ & 
\makecell[{{p{5cm}}}]{\raggedright$f_{1}=11224343244240133023443242, n_1=25$ \\$f_{2}=120201131241411221034314300400010003010$\\$413211341111, n_2=50$} & 
\makecell[{{p{5cm}}}]{\raggedright$g_{1}=122131442144013344000013$ \\$g_{2}=1102114342343003301434$\\$3333103030434014041100340442$} & 
\makecell[{{p{5cm}}}]{\raggedright$p_{1}=20$ \\$p_{2}=44$} & 
$[75,2,60]_7$ & $[[75,69,2]]_5$& \cite{dinh2020}\\

\hline
$[[84,78,2]]_3$ & 
\makecell[{{p{5cm}}}]{\raggedright$f_{1}=100220222212012$\\$2100200021200222002020, n_1=36$ \\$f_{2}=10001102001001012002020000$\\$2110211021021211000012, n_2=50$} & 
\makecell[{{p{5cm}}}]{\raggedright$g_{1}=101201001$\\$1212211112112202220201112$ \\$g_{2}=10210221222200021111$\\$12111012120011011021000200$} & 
\makecell[{{p{5cm}}}]{\raggedright$p_{1}=211$ \\$p_{2}=211$} & 
$[84,3,51]_3$ & $[[84,78,2]]_3$& \cite{dinh2020} \\

\hline

$[[90,86,2]]_5*$ & 
\makecell[{{p{5cm}}}]{\raggedright$f_{1}=13314302020014200243124231031$\\$112224301131101142414014, n_1=52$ \\$f_{2}=102101013003140401432, n_2=20$
\\$f_{3}=122140211421423003, n_3=18$}& 
\makecell[{{p{5cm}}}]{\raggedright$g_{1}=102100441022331044020301221$\\$311410214040424433414440$ \\$g_{2}=1422312204042331414$ \\ $g_{3}=13200142314144321$} & 
\makecell[{{p{5cm}}}]{\raggedright$p_{1}=21=$ \\$p_{2}=2=$\\$p_{3}=41$} & 
$[90,2,71]_5$ & $[[90,84,2]]_5$& \cite{dinh2020-2} \\

\hline

$[[100,96,2]]_5*$ & 
\makecell[{{p{5cm}}}]{\raggedright$f_{1}=13314302020014200243124231031$\\$112224301131101142414014, n_1=52$ \\$f_{2}=102101013003140401432, n_2=20$
\\$f_{3}=122140211421423003, n_3=18$}& 
\makecell[{{p{5cm}}}]{\raggedright$g_{1}=102100441022331044020301221$\\$311410214040424433414440$ \\$g_{2}=1422312204042331414$ \\ $g_{3}=13200142314144321$} & 
\makecell[{{p{5cm}}}]{\raggedright$p_{1}=21=$ \\$p_{2}=2=$\\$p_{3}=41$} & 
$[90,2,71]_5$ & $[[100,94,2]]_5$& \cite{dinh2020-2} \\

\hline
\end{tabular*}
\end{minipage}
\end{center}
*: better code compared to the codes in references.
\end{table}

\restoregeometry

\newgeometry{left=3.5cm, right=3.5cm, top=2.4cm, bottom=1.9cm}

\section{Acknowledgements}
We thank the Kenyon Summer Science Scholars program for providing funding for this research. 

\end{document}